
%
%
%
%
%
%
%
\def\standardrisposta{s }\def\reducedrisposta{r }
\def\mplarisposta{mpla }
\def\doublerisposta{d }\def\cartarisposta{e }\def\amsrisposta{y }
\newcount\ingrandimento \newcount\sinnota \newcount\dimnota
\newcount\unoduecol \newdimen\collhsize \newdimen\tothsize
\newdimen\fullhsize \newcount\controllorisposta \sinnota=1
\newskip\infralinea  \global\controllorisposta=0
\message{ ********    Welcome to PANDA macros (Plain TeX, AP, 1991)}
\message{ ******** }
\message{       You'll have to answer a few questions in lowercase.}
\message{>  Do you want it in double-page (d), reduced (r)}
\message{or standard format (s) ? }\read-1 to\risposta
\message{>  Do you want it in USA A4 (u) or EUROPEAN A4 (e)}
\message{paper size ? }\read-1 to\srisposta
\message{>  Do you have AMSFonts 2.0 (math) fonts (y/n) ? }
\read-1 to\arisposta
%
%
%
%
%
\ifx\risposta\standardrisposta \ingrandimento=1200
\message{>> This will come out UNREDUCED << }
\dimnota=2 \unoduecol=1 \global\controllorisposta=1 \fi
\ifx\risposta\reducedrisposta \ingrandimento=1095 \dimnota=1
\unoduecol=1  \global\controllorisposta=1
\message{>> This will come out REDUCED << } \fi
\ifx\risposta\doublerisposta \ingrandimento=1000 \dimnota=2
\unoduecol=2  \global\controllorisposta=1 
\message{>> You must print this in LANDSCAPE orientation << } \fi
\ifx\risposta\mplarisposta \ingrandimento=1000 \dimnota=1
\message{>> Mod. Phys. Lett. A format << }
\unoduecol=1 \global\controllorisposta=1 \fi
\ifnum\controllorisposta=0  \ingrandimento=1200
\message{>>> ERROR IN INPUT, I ASSUME STANDARD UNREDUCED FORMAT <<< }
\dimnota=2 \unoduecol=1 \fi
\magnification=\ingrandimento
%
%
%
%
\newdimen\eucolumnsize \newdimen\eudoublehsize \newdimen\eudoublevsize
\newdimen\uscolumnsize \newdimen\usdoublehsize \newdimen\usdoublevsize
\newdimen\eusinglehsize \newdimen\eusinglevsize \newdimen\ussinglehsize
\newskip\standardbaselineskip \newdimen\ussinglevsize
\newskip\reducedbaselineskip \newskip\doublebaselineskip
\eucolumnsize=12.0truecm    
\eudoublehsize=25.5truecm   
\eudoublevsize=6.5truein    
\uscolumnsize=4.4truein     
\usdoublehsize=9.4truein    
\usdoublevsize=6.8truein    
\eusinglehsize=6.5truein    
\eusinglevsize=24truecm     
\ussinglehsize=6.5truein    
\ussinglevsize=8.9truein    
\standardbaselineskip=16pt  
\reducedbaselineskip=14pt   
\doublebaselineskip=12pt    
%
%
\def\Portoffset{}
\def\Landoffset{}
\ifx\risposta\mplarisposta \def\Portoffset{\hoffset=1.8truecm} \fi
%
%
\def\Landspec{}
\tolerance=10000
\parskip 0pt plus 2pt  \leftskip=0pt \rightskip=0pt
%
%
\ifx\risposta\standardrisposta \infralinea=\standardbaselineskip \fi
\ifx\risposta\reducedrisposta  \infralinea=\reducedbaselineskip \fi
\ifx\risposta\doublerisposta   \infralinea=\doublebaselineskip \fi
\ifx\risposta\mplarisposta     \infralinea=13pt \fi
\ifnum\controllorisposta=0    \infralinea=\standardbaselineskip \fi
\ifx\risposta\doublerisposta   \Landoffset \else \Portoffset \fi
\ifx\risposta\doublerisposta \ifx\srisposta\cartarisposta
\tothsize=\eudoublehsize \collhsize=\eucolumnsize
\vsize=\eudoublevsize  \else  \tothsize=\usdoublehsize
\collhsize=\uscolumnsize \vsize=\usdoublevsize \fi \else
\ifx\srisposta\cartarisposta \tothsize=\eusinglehsize
\vsize=\eusinglevsize \else  \tothsize=\ussinglehsize
\vsize=\ussinglevsize \fi \collhsize=4.4truein \fi
\ifx\risposta\mplarisposta \tothsize=5.0truein
\vsize=7.8truein \collhsize=4.4truein \fi
%
%
%
%
\newcount\contaeuler \newcount\contacyrill \newcount\contaams
\font\ninerm=cmr9  \font\eightrm=cmr8  \font\sixrm=cmr6
\font\ninei=cmmi9  \font\eighti=cmmi8  \font\sixi=cmmi6
\font\ninesy=cmsy9  \font\eightsy=cmsy8  \font\sixsy=cmsy6
\font\ninebf=cmbx9  \font\eightbf=cmbx8  \font\sixbf=cmbx6
\font\ninett=cmtt9  \font\eighttt=cmtt8  \font\nineit=cmti9
\font\eightit=cmti8 \font\ninesl=cmsl9  \font\eightsl=cmsl8
\skewchar\ninei='177 \skewchar\eighti='177 \skewchar\sixi='177
\skewchar\ninesy='60 \skewchar\eightsy='60 \skewchar\sixsy='60
\hyphenchar\ninett=-1 \hyphenchar\eighttt=-1 \hyphenchar\tentt=-1
%
\font\tencmmib=cmmib10  \newfam\cmmibfam  \skewchar\tencmmib='177
\font\tencmbsy=cmbsy10  \newfam\cmbsyfam  \skewchar\tencmbsy='60
\def\scaps{\cmcsc}                 
\font\tencmcsc=cmcsc10  \newfam\cmcscfam
\ifnum\ingrandimento=1095

\font\capsone=cmcsc10 at 10.95pt 

\else

\font\capsone=cmcsc10 at 12pt 
\fi

\def\ttaarr{\bf}		
\def\ppaarr{\sl}		

%
%
%
\newfam\eufmfam \newfam\msamfam \newfam\msbmfam \newfam\eufbfam
\def\Loadeulerfonts{\global\contaeuler=1 \ifx\arisposta\amsrisposta
\font\teneufm=eufm10              
\font\eighteufm=eufm8 \font\nineeufm=eufm9 \font\sixeufm=eufm6
\font\seveneufm=eufm7  \font\fiveeufm=eufm5
\font\teneufb=eufb10              
\font\eighteufb=eufb8 \font\nineeufb=eufb9 \font\sixeufb=eufb6
\font\seveneufb=eufb7  \font\fiveeufb=eufb5
\font\teneurm=eurm10              
\font\eighteurm=eurm8 \font\nineeurm=eurm9
\font\teneurb=eurb10              
\font\eighteurb=eurb8 \font\nineeurb=eurb9
\font\teneusm=eusm10              
\font\eighteusm=eusm8 \font\nineeusm=eusm9
\font\teneusb=eusb10              
\font\eighteusb=eusb8 \font\nineeusb=eusb9
\else \def\eufm{\tt} \def\eufb{\tt} \def\eurm{\tt} \def\eurb{\tt}
\def\eusm{\tt} \def\eusb{\tt}    \fi}

\def\loadamsmath{\global\contaams=1 \ifx\arisposta\amsrisposta
\font\tenmsam=msam10 \font\ninemsam=msam9 \font\eightmsam=msam8
\font\sevenmsam=msam7 \font\sixmsam=msam6 \font\fivemsam=msam5
\font\tenmsbm=msbm10 \font\ninemsbm=msbm9 \font\eightmsbm=msbm8
\font\sevenmsbm=msbm7 \font\sixmsbm=msbm6 \font\fivemsbm=msbm5
\else \def\msbm{\bf} \fi \def\Bbb{\msbm} \def\symbl{\msam} \tenpoint}
\def\loadcyrill{\global\contacyrill=1 \ifx\arisposta\amsrisposta
\font\tenwncyr=wncyr10 \font\ninewncyr=wncyr9 \font\eightwncyr=wncyr8
\font\tenwncyb=wncyr10 \font\ninewncyb=wncyr9 \font\eightwncyb=wncyr8
\font\tenwncyi=wncyr10 \font\ninewncyi=wncyr9 \font\eightwncyi=wncyr8
\else \def\cyrill{\sl} \def\cyrilb{\sl} \def\cyrili{\sl} \fi\tenpoint}
\ifx\arisposta\amsrisposta
\font\sevenex=cmex7               
\font\eightex=cmex8  \font\nineex=cmex9
\font\ninecmmib=cmmib9   \font\eightcmmib=cmmib8
\font\sevencmmib=cmmib7 \font\sixcmmib=cmmib6
\font\fivecmmib=cmmib5   \skewchar\ninecmmib='177
\skewchar\eightcmmib='177  \skewchar\sevencmmib='177
\skewchar\sixcmmib='177   \skewchar\fivecmmib='177
\font\ninecmbsy=cmbsy9    \font\eightcmbsy=cmbsy8
\font\sevencmbsy=cmbsy7  \font\sixcmbsy=cmbsy6
\font\fivecmbsy=cmbsy5   \skewchar\ninecmbsy='60
\skewchar\eightcmbsy='60  \skewchar\sevencmbsy='60
\skewchar\sixcmbsy='60    \skewchar\fivecmbsy='60
\font\ninecmcsc=cmcsc9    \font\eightcmcsc=cmcsc8     \else
\def\cmmib{\fam\cmmibfam\tencmmib}\textfont\cmmibfam=\tencmmib
\scriptfont\cmmibfam=\tencmmib \scriptscriptfont\cmmibfam=\tencmmib
\def\cmbsy{\fam\cmbsyfam\tencmbsy} \textfont\cmbsyfam=\tencmbsy
\scriptfont\cmbsyfam=\tencmbsy \scriptscriptfont\cmbsyfam=\tencmbsy
\scriptfont\cmcscfam=\tencmcsc \scriptscriptfont\cmcscfam=\tencmcsc
\def\cmcsc{\fam\cmcscfam\tencmcsc} \textfont\cmcscfam=\tencmcsc \fi
\catcode`@=11
\newskip\ttglue
\gdef\tenpoint{\def\rm{\fam0\tenrm}
  \textfont0=\tenrm \scriptfont0=\sevenrm \scriptscriptfont0=\fiverm
  \textfont1=\teni \scriptfont1=\seveni \scriptscriptfont1=\fivei
  \textfont2=\tensy \scriptfont2=\sevensy \scriptscriptfont2=\fivesy
  \textfont3=\tenex \scriptfont3=\tenex \scriptscriptfont3=\tenex
  \def\mcal{\fam2 \tensy}  \def\mmit{\fam1 \teni}
  \textfont\itfam=\tenit \def\it{\fam\itfam\tenit}
  \textfont\slfam=\tensl \def\sl{\fam\slfam\tensl}
  \textfont\ttfam=\tentt \scriptfont\ttfam=\eighttt
  \scriptscriptfont\ttfam=\eighttt  \def\tt{\fam\ttfam\tentt}
  \textfont\bffam=\tenbf \scriptfont\bffam=\sevenbf
  \scriptscriptfont\bffam=\fivebf \def\bf{\fam\bffam\tenbf}
     \ifx\arisposta\amsrisposta    \ifnum\contaeuler=1
  \textfont\eufmfam=\teneufm \scriptfont\eufmfam=\seveneufm
  \scriptscriptfont\eufmfam=\fiveeufm \def\eufm{\fam\eufmfam\teneufm}
  \textfont\eufbfam=\teneufb \scriptfont\eufbfam=\seveneufb
  \scriptscriptfont\eufbfam=\fiveeufb \def\eufb{\fam\eufbfam\teneufb}
  \def\eurm{\teneurm} \def\eurb{\teneurb} \def\eusm{\teneusm}
  \def\eusb{\teneusb}    \fi    \ifnum\contaams=1
  \textfont\msamfam=\tenmsam \scriptfont\msamfam=\sevenmsam
  \scriptscriptfont\msamfam=\fivemsam \def\msam{\fam\msamfam\tenmsam}
  \textfont\msbmfam=\tenmsbm \scriptfont\msbmfam=\sevenmsbm
  \scriptscriptfont\msbmfam=\fivemsbm \def\msbm{\fam\msbmfam\tenmsbm}
     \fi      \ifnum\contacyrill=1     \def\cyrill{\tenwncyr}
  \def\cyrilb{\tenwncyb}  \def\cyrili{\tenwncyi}         \fi
  \textfont3=\tenex \scriptfont3=\sevenex \scriptscriptfont3=\sevenex
  \def\cmmib{\fam\cmmibfam\tencmmib} \scriptfont\cmmibfam=\sevencmmib
  \textfont\cmmibfam=\tencmmib  \scriptscriptfont\cmmibfam=\fivecmmib
  \def\cmbsy{\fam\cmbsyfam\tencmbsy} \scriptfont\cmbsyfam=\sevencmbsy
  \textfont\cmbsyfam=\tencmbsy  \scriptscriptfont\cmbsyfam=\fivecmbsy
  \def\cmcsc{\fam\cmcscfam\tencmcsc} \scriptfont\cmcscfam=\eightcmcsc
  \textfont\cmcscfam=\tencmcsc \scriptscriptfont\cmcscfam=\eightcmcsc
     \fi            \tt \ttglue=.5em plus.25em minus.15em
  \normalbaselineskip=12pt
  \setbox\strutbox=\hbox{\vrule height8.5pt depth3.5pt width0pt}
  \let\sc=\eightrm \let\big=\tenbig   \normalbaselines
  \baselineskip=\infralinea  \rm}
\gdef\ninepoint{\def\rm{\fam0\ninerm}
  \textfont0=\ninerm \scriptfont0=\sixrm \scriptscriptfont0=\fiverm
  \textfont1=\ninei \scriptfont1=\sixi \scriptscriptfont1=\fivei
  \textfont2=\ninesy \scriptfont2=\sixsy \scriptscriptfont2=\fivesy
  \textfont3=\tenex \scriptfont3=\tenex \scriptscriptfont3=\tenex
  \def\mcal{\fam2 \ninesy}  \def\mmit{\fam1 \ninei}
  \textfont\itfam=\nineit \def\it{\fam\itfam\nineit}
  \textfont\slfam=\ninesl \def\sl{\fam\slfam\ninesl}
  \textfont\ttfam=\ninett \scriptfont\ttfam=\eighttt
  \scriptscriptfont\ttfam=\eighttt \def\tt{\fam\ttfam\ninett}
  \textfont\bffam=\ninebf \scriptfont\bffam=\sixbf
  \scriptscriptfont\bffam=\fivebf \def\bf{\fam\bffam\ninebf}
     \ifx\arisposta\amsrisposta  \ifnum\contaeuler=1
  \textfont\eufmfam=\nineeufm \scriptfont\eufmfam=\sixeufm
  \scriptscriptfont\eufmfam=\fiveeufm \def\eufm{\fam\eufmfam\nineeufm}
  \textfont\eufbfam=\nineeufb \scriptfont\eufbfam=\sixeufb
  \scriptscriptfont\eufbfam=\fiveeufb \def\eufb{\fam\eufbfam\nineeufb}
  \def\eurm{\nineeurm} \def\eurb{\nineeurb} \def\eusm{\nineeusm}
  \def\eusb{\nineeusb}     \fi   \ifnum\contaams=1
  \textfont\msamfam=\ninemsam \scriptfont\msamfam=\sixmsam
  \scriptscriptfont\msamfam=\fivemsam \def\msam{\fam\msamfam\ninemsam}
  \textfont\msbmfam=\ninemsbm \scriptfont\msbmfam=\sixmsbm
  \scriptscriptfont\msbmfam=\fivemsbm \def\msbm{\fam\msbmfam\ninemsbm}
     \fi       \ifnum\contacyrill=1     \def\cyrill{\ninewncyr}
  \def\cyrilb{\ninewncyb}  \def\cyrili{\ninewncyi}         \fi
  \textfont3=\nineex \scriptfont3=\sevenex \scriptscriptfont3=\sevenex
  \def\cmmib{\fam\cmmibfam\ninecmmib}  \textfont\cmmibfam=\ninecmmib
  \scriptfont\cmmibfam=\sixcmmib \scriptscriptfont\cmmibfam=\fivecmmib
  \def\cmbsy{\fam\cmbsyfam\ninecmbsy}  \textfont\cmbsyfam=\ninecmbsy
  \scriptfont\cmbsyfam=\sixcmbsy \scriptscriptfont\cmbsyfam=\fivecmbsy
  \def\cmcsc{\fam\cmcscfam\ninecmcsc} \scriptfont\cmcscfam=\eightcmcsc
  \textfont\cmcscfam=\ninecmcsc \scriptscriptfont\cmcscfam=\eightcmcsc
     \fi            \tt \ttglue=.5em plus.25em minus.15em
  \normalbaselineskip=11pt
  \setbox\strutbox=\hbox{\vrule height8pt depth3pt width0pt}
  \let\sc=\sevenrm \let\big=\ninebig \normalbaselines\rm}
\gdef\eightpoint{\def\rm{\fam0\eightrm}
  \textfont0=\eightrm \scriptfont0=\sixrm \scriptscriptfont0=\fiverm
  \textfont1=\eighti \scriptfont1=\sixi \scriptscriptfont1=\fivei
  \textfont2=\eightsy \scriptfont2=\sixsy \scriptscriptfont2=\fivesy
  \textfont3=\tenex \scriptfont3=\tenex \scriptscriptfont3=\tenex
  \def\mcal{\fam2 \eightsy}  \def\mmit{\fam1 \eighti}
  \textfont\itfam=\eightit \def\it{\fam\itfam\eightit}
  \textfont\slfam=\eightsl \def\sl{\fam\slfam\eightsl}
  \textfont\ttfam=\eighttt \scriptfont\ttfam=\eighttt
  \scriptscriptfont\ttfam=\eighttt \def\tt{\fam\ttfam\eighttt}
  \textfont\bffam=\eightbf \scriptfont\bffam=\sixbf
  \scriptscriptfont\bffam=\fivebf \def\bf{\fam\bffam\eightbf}
     \ifx\arisposta\amsrisposta   \ifnum\contaeuler=1
  \textfont\eufmfam=\eighteufm \scriptfont\eufmfam=\sixeufm
  \scriptscriptfont\eufmfam=\fiveeufm \def\eufm{\fam\eufmfam\eighteufm}
  \textfont\eufbfam=\eighteufb \scriptfont\eufbfam=\sixeufb
  \scriptscriptfont\eufbfam=\fiveeufb \def\eufb{\fam\eufbfam\eighteufb}
  \def\eurm{\eighteurm} \def\eurb{\eighteurb} \def\eusm{\eighteusm}
  \def\eusb{\eighteusb}       \fi    \ifnum\contaams=1
  \textfont\msamfam=\eightmsam \scriptfont\msamfam=\sixmsam
  \scriptscriptfont\msamfam=\fivemsam \def\msam{\fam\msamfam\eightmsam}
  \textfont\msbmfam=\eightmsbm \scriptfont\msbmfam=\sixmsbm
  \scriptscriptfont\msbmfam=\fivemsbm \def\msbm{\fam\msbmfam\eightmsbm}
     \fi       \ifnum\contacyrill=1     \def\cyrill{\eightwncyr}
  \def\cyrilb{\eightwncyb}  \def\cyrili{\eightwncyi}         \fi
  \textfont3=\eightex \scriptfont3=\sevenex \scriptscriptfont3=\sevenex
  \def\cmmib{\fam\cmmibfam\eightcmmib}  \textfont\cmmibfam=\eightcmmib
  \scriptfont\cmmibfam=\sixcmmib \scriptscriptfont\cmmibfam=\fivecmmib
  \def\cmbsy{\fam\cmbsyfam\eightcmbsy}  \textfont\cmbsyfam=\eightcmbsy
  \scriptfont\cmbsyfam=\sixcmbsy \scriptscriptfont\cmbsyfam=\fivecmbsy
  \def\cmcsc{\fam\cmcscfam\eightcmcsc} \scriptfont\cmcscfam=\eightcmcsc
  \textfont\cmcscfam=\eightcmcsc \scriptscriptfont\cmcscfam=\eightcmcsc
     \fi             \tt \ttglue=.5em plus.25em minus.15em
  \normalbaselineskip=9pt
  \setbox\strutbox=\hbox{\vrule height7pt depth2pt width0pt}
  \let\sc=\sixrm \let\big=\eightbig \normalbaselines\rm }
\gdef\tenbig#1{{\hbox{$\left#1\vbox to8.5pt{}\right.\n@space$}}}
\gdef\ninebig#1{{\hbox{$\textfont0=\tenrm\textfont2=\tensy
   \left#1\vbox to7.25pt{}\right.\n@space$}}}
\gdef\eightbig#1{{\hbox{$\textfont0=\ninerm\textfont2=\ninesy
   \left#1\vbox to6.5pt{}\right.\n@space$}}}
\def\alternativefont#1#2{\ifx\arisposta\amsrisposta \relax \else
\xdef#1{#2} \fi}
\global\contaeuler=0 \global\contacyrill=0 \global\contaams=0
%
%
%
%
\newbox\fotlinebb \newbox\hedlinebb \newbox\leftcolumn
\gdef\makeheadline{\vbox to 0pt{\vskip-22.5pt
     \fullline{\vbox to8.5pt{}\the\headline}\vss}\nointerlineskip}
\gdef\makehedlinebb{\vbox to 0pt{\vskip-22.5pt
     \fullline{\vbox to8.5pt{}\copy\hedlinebb\hfil
     \line{\hfill\the\headline\hfill}}\vss} \nointerlineskip}
\gdef\makefootline{\baselineskip=24pt \fullline{\the\footline}}
\gdef\makefotlinebb{\baselineskip=24pt
    \fullline{\copy\fotlinebb\hfil\line{\hfill\the\footline\hfill}}}
\gdef\doubleformat{\shipout\vbox{\Landspec\makehedlinebb
     \fullline{\box\leftcolumn\hfil\columnbox}\makefotlinebb}
     \advancepageno}
\gdef\columnbox{\leftline{\pagebody}}
\gdef\line#1{\hbox to\hsize{\hskip\leftskip#1\hskip\rightskip}}
\gdef\fullline#1{\hbox to\fullhsize{\hskip\leftskip{#1}%
\hskip\rightskip}}
\gdef\footnote#1{\let\@sf=\empty
         \ifhmode\edef\#sf{\spacefactor=\the\spacefactor}\/\fi
         #1\@sf\vfootnote{#1}}
\gdef\vfootnote#1{\insert\footins\bgroup
         \ifnum\dimnota=1  \eightpoint\fi
         \ifnum\dimnota=2  \ninepoint\fi
         \ifnum\dimnota=0  \tenpoint\fi
         \interlinepenalty=\interfootnotelinepenalty
         \splittopskip=\ht\strutbox
         \splitmaxdepth=\dp\strutbox \floatingpenalty=20000
         \leftskip=\oldssposta \rightskip=\olddsposta
         \spaceskip=0pt \xspaceskip=0pt
         \ifnum\sinnota=0   \textindent{#1}\fi
         \ifnum\sinnota=1   \item{#1}\fi
         \footstrut\futurelet\next\fo@t}
\gdef\fo@t{\ifcat\bgroup\noexpand\next \let\next\f@@t
             \else\let\next\f@t\fi \next}
\gdef\f@@t{\bgroup\aftergroup\@foot\let\next}
\gdef\f@t#1{#1\@foot} \gdef\@foot{\strut\egroup}
\gdef\footstrut{\vbox to\splittopskip{}}
\skip\footins=\bigskipamount
\count\footins=1000  \dimen\footins=8in
\catcode`@=12
\tenpoint
\ifnum\unoduecol=1 \hsize=\tothsize   \fullhsize=\tothsize \fi
\ifnum\unoduecol=2 \hsize=\collhsize  \fullhsize=\tothsize \fi
\global\let\lrcol=L
\ifnum\unoduecol=1 \output{\plainoutput{\ifnum\tipbnota=2
\clearnmbnota\fi}} \fi
\ifnum\unoduecol=2 \output{\if L\lrcol
     \global\setbox\leftcolumn=\columnbox
     \global\setbox\fotlinebb=\line{\hfill\the\footline\hfill}
     \global\setbox\hedlinebb=\line{\hfill\the\headline\hfill}
     \advancepageno  \global\let\lrcol=R
     \else  \doubleformat \global\let\lrcol=L \fi
     \ifnum\outputpenalty>-20000 \else\dosupereject\fi
     \ifnum\tipbnota=2\clearnmbnota\fi }\fi
\def\ifdoublepage{\ifnum\unoduecol=2 }
\gdef\yespagenumbers{\footline={\hss\tenrm\folio\hss}}
\gdef\ciao{\par\vfill\supereject \ifnum\unoduecol=2
     \if R\lrcol  \headline={}\nopagenumbers\null\vfill\eject
     \fi\fi \end}

\newskip\olddsposta \newskip\oldssposta
\global\oldssposta=\leftskip \global\olddsposta=\rightskip

\def\filldots{\leaders\hbox to 1em{\hss.\hss}\hfill}
\def\inquadrb#1 {\vbox {\hrule  \hbox{\vrule \vbox {\vskip .2cm
    \hbox {\ #1\ } \vskip .2cm } \vrule  }  \hrule} }
 \def\newline{\hfil\break}
\def\jump{\vskip\baselineskip} \newskip\iinnffrr
\def\sjump{\iinnffrr=\baselineskip
          \divide\iinnffrr by 2 \vskip\iinnffrr}
\def\bjump{\vskip\baselineskip \vskip\baselineskip}
\newcount\nmbnota  \def\clearnmbnota{\global\nmbnota=0}
\newcount\tipbnota \def\letterfootnote{\global\tipbnota=1}

\def\note#1{\global\advance\nmbnota by 1 \ifnum\tipbnota=1
    \footnote{$^{\rm\nttlett}$}{#1} \else {\ifnum\tipbnota=2
    \footnote{$^{\nttsymb}$}{#1}
    \else\footnote{$^{\the\nmbnota}$}{#1}\fi}\fi}
\def\nttlett{\ifcase\nmbnota \or a\or b\or c\or d\or e\or f\or
g\or h\or i\or j\or k\or l\or m\or n\or o\or p\or q\or r\or
s\or t\or u\or v\or w\or y\or x\or z\fi}
\def\nttsymb{\ifcase\nmbnota \or\dag\or\sharp\or\ddag\or\star\or
\natural\or\flat\or\clubsuit\or\diamondsuit\or\heartsuit
\or\spadesuit\fi}   \clearnmbnota
\def\numberfootnote{\global\tipbnota=0} \numberfootnote
\def\setnote#1{\expandafter\xdef\csname#1\endcsname{
\ifnum\tipbnota=1 {\rm\nttlett} \else {\ifnum\tipbnota=2
{\nttsymb} \else \the\nmbnota\fi}\fi} }
\newcount\nbmfig  \def\clearnbmfig{\global\nbmfig=0}
\gdef\figure{\global\advance\nbmfig by 1
      {\rm fig. \the\nbmfig}}   \clearnbmfig
\def\setfig#1{\expandafter\xdef\csname#1\endcsname{fig. \the\nbmfig}}

\newcount\frmcount \def\clearfrmcount{\global\frmcount=0}
\def\numero{\global\advance\frmcount by 1   \ifnum\indappcount=0
  {\ifnum\cpcount <1 {\hbox{\rm (\the\frmcount )}}  \else
  {\hbox{\rm (\the\cpcount .\the\frmcount )}} \fi}  \else
  {\hbox{\rm (\applett .\the\frmcount )}} \fi}
\def\nameformula#1{\global\advance\frmcount by 1%
\ifnum\draftnum=0  {\ifnum\indappcount=0%
{\ifnum\cpcount<1\xdef\spzzttrra{(\the\frmcount )}%
\else\xdef\spzzttrra{(\the\cpcount .\the\frmcount )}\fi}%
\else\xdef\spzzttrra{(\applett .\the\frmcount )}\fi}%
\else\xdef\spzzttrra{(#1)}\fi%
\expandafter\xdef\csname#1\endcsname{\spzzttrra}
\eqno \ifnum\draftnum=0 {\ifnum\indappcount=0
  {\ifnum\cpcount <1 {\hbox{\rm (\the\frmcount )}}  \else
  {\hbox{\rm (\the\cpcount .\the\frmcount )}}\fi}   \else
  {\hbox{\rm (\applett .\the\frmcount )}} \fi} \else (#1) \fi $$}
\def\nfr{\nameformula}    
\def\nameali#1{\global\advance\frmcount by 1%
\ifnum\draftnum=0  {\ifnum\indappcount=0%
{\ifnum\cpcount<1\xdef\spzzttrra{(\the\frmcount )}%
\else\xdef\spzzttrra{(\the\cpcount .\the\frmcount )}\fi}%
\else\xdef\spzzttrra{(\applett .\the\frmcount )}\fi}%
\else\xdef\spzzttrra{(#1)}\fi%
\expandafter\xdef\csname#1\endcsname{\spzzttrra}
  \ifnum\draftnum=0  {\ifnum\indappcount=0
  {\ifnum\cpcount <1 {\hbox{\rm (\the\frmcount )}}  \else
  {\hbox{\rm (\the\cpcount .\the\frmcount )}}\fi}   \else
  {\hbox{\rm (\applett .\the\frmcount )}} \fi} \else (#1) \fi}
\clearfrmcount
\newcount\cpcount \def\clearcpcount{\global\cpcount=0}
\newcount\subcpcount \def\clearsubcpcount{\global\subcpcount=0}
\newcount\appcount \def\clearappcount{\global\appcount=0}
\newcount\indappcount \def\clearindappcount{\indappcount=0}
\newcount\sottoparcount 

\def\applett{\ifcase\appcount  \or {A}\or {B}\or {C}\or
{D}\or {E}\or {F}\or {G}\or {H}\or {I}\or {J}\or {K}\or {L}\or
{M}\or {N}\or {O}\or {P}\or {Q}\or {R}\or {S}\or {T}\or {U}\or
{V}\or {W}\or {X}\or {Y}\or {Z}\fi
             \ifnum\appcount<0
    \message{>>  ERROR: counter \appcount out of range <<}\fi
             \ifnum\appcount>26
   \message{>>  ERROR: counter \appcount out of range <<}\fi}
\clearappcount  \clearindappcount
\newcount\connttrre  \def\clearconnttrre{\global\connttrre=0}
\newcount\countref  \def\clearcountref{\global\countref=0}
\clearcountref
\def\chapter#1{\global\advance\cpcount by 1 \clearfrmcount
                 \goodbreak\null\vbox{\jump\nobreak
                 \clearsubcpcount\clearindappcount
                 \itemitem{\ttaarr\the\cpcount .\qquad}{\ttaarr #1}
                 \par\nobreak\jump\sjump}\nobreak}
\def\section#1{\global\advance\subcpcount by 1 \goodbreak\null
               \vbox{\sjump\nobreak\ifnum\indappcount=0
                 {\ifnum\cpcount=0 {\itemitem{\ppaarr
               .\the\subcpcount\quad\enskip\ }{\ppaarr #1}\par} \else
                 {\itemitem{\ppaarr\the\cpcount .\the\subcpcount\quad
                  \enskip\ }{\ppaarr #1} \par}  \fi}
                \else{\itemitem{\ppaarr\applett .\the\subcpcount\quad
                 \enskip\ }{\ppaarr #1}\par}\fi\nobreak\jump}\nobreak}
\clearsubcpcount
\def\appendix#1{\global\advance\appcount by 1 \clearfrmcount
                  \goodbreak\null\vbox{\jump\nobreak
                  \global\advance\indappcount by 1 \clearsubcpcount
                  \itemitem{\ttaarr App.\applett\ }{\ttaarr #1}
                  \nobreak\jump\sjump}\nobreak}
\clearappcount \clearindappcount

\clearcpcount\clearcountref

\def\setchap#1{\ifnum\indappcount=0{\ifnum\subcpcount=0%
\xdef\spzzttrra{\the\cpcount}%
\else\xdef\spzzttrra{\the\cpcount .\the\subcpcount}\fi}
\else{\ifnum\subcpcount=0 \xdef\spzzttrra{\applett}%
\else\xdef\spzzttrra{\applett .\the\subcpcount}\fi}\fi
\expandafter\xdef\csname#1\endcsname{\spzzttrra}}
\newcount\draftnum \newcount\ppora   \newcount\ppminuti
\global\ppora=\time   \global\ppminuti=\time
\global\divide\ppora by 60  \draftnum=\ppora
\multiply\draftnum by 60    \global\advance\ppminuti by -\draftnum
\global\draftnum=0
\def\droggi{\number\day /\number\month /\number\year\ \the\ppora
:\the\ppminuti}
 \global\draftnum=0
\def\draftcomment#1{\ifnum\draftnum=0 \relax \else
{\ {\bf ***}\ #1\ {\bf ***}\ }\fi} 
%
%
\catcode`@=11
\gdef\Ref#1{\expandafter\ifx\csname @rrxx@#1\endcsname\relax%
{\global\advance\countref by 1%
\ifnum\countref>200%
\message{>>> ERROR: maximum number of references exceeded <<<}%
\expandafter\xdef\csname @rrxx@#1\endcsname{0}\else%
\expandafter\xdef\csname @rrxx@#1\endcsname{\the\countref}\fi}\fi%
\ifnum\draftnum=0 \csname @rrxx@#1\endcsname \else#1\fi}
\gdef\beginref{\ifnum\draftnum=0  \gdef\Rref{\fairef}
\gdef\endref{\scriviref} \else\relax\fi
\ifx\risposta\mplarisposta \ninepoint \fi
\parskip 2pt plus 2pt \baselineskip=12pt}
\def\Reflab#1{[#1]} \gdef\Rref#1#2{\item{\Reflab{#1}}{#2}}
\gdef\endref{\relax}  \newcount\conttemp
\gdef\fairef#1#2{\expandafter\ifx\csname @rrxx@#1\endcsname\relax
{\global\conttemp=0
\message{>>> ERROR: reference [#1] not defined <<<} } \else
{\global\conttemp=\csname @rrxx@#1\endcsname } \fi
\global\advance\conttemp by 50
\global\setbox\conttemp=\hbox{#2} }
\gdef\scriviref{\clearconnttrre\conttemp=50
\loop\ifnum\connttrre<\countref \advance\conttemp by 1
\advance\connttrre by 1
\item{\Reflab{\the\connttrre}}{\unhcopy\conttemp} \repeat}
\clearcountref \clearconnttrre
\catcode`@=12
\ifx\risposta\mplarisposta \def\Reflab#1{#1.} \letterfootnote \fi

\def\slashchar#1{\setbox0=\hbox{$#1$} \dimen0=\wd0
     \setbox1=\hbox{/} \dimen1=\wd1 \ifdim\dimen0>\dimen1
      \rlap{\hbox to \dimen0{\hfil/\hfil}} #1 \else
      \rlap{\hbox to \dimen1{\hfil$#1$\hfil}} / \fi}
\ifx\oldchi\undefined \let\oldchi=\chi
  \def\cchi{{\raise 1pt\hbox{$\oldchi$}}} \let\chi=\cchi \fi
  
\def\del{\partial}   

\def\frac#1#2{{\textstyle{#1 \over #2}}}

\def\half{\ifinner {\scriptstyle {1 \over 2}}\else {1 \over 2} \fi}

\def\simge{\rlap{\raise 2pt \hbox{$>$}}{\lower 2pt \hbox{$\sim$}}}
\def\simle{\rlap{\raise 2pt \hbox{$<$}}{\lower 2pt \hbox{$\sim$}}}

\def\vbig#1#2{{\vbigd@men=#2\divide\vbigd@men by 2%
\hbox{$\left#1\vbox to \vbigd@men{}\right.\n@space$}}}

\null
%
%
%
%



\loadamsmath
\font\mtitle=cmmi10 at 12pt

\def\del{\partial}
\def\frac#1#2{ {#1 \over #2} }

\def\A{ A }
\def\B{ B }
\def\C{ C }

\def\a{\alpha}     \def\ad{{\dot \alpha}}
\def\b{\beta}      \def\bd{{\dot \beta}}
     \def\G{\Gamma}
\def\d{\delta}     
\def\e{\epsilon}   \def\ve{\varepsilon}
\def\eh{{\hat \e}} \def\eb{{\bar \e}}
\def\t{\tau}
\def\vh{{\hat v}}   \def\vb{{\bar v}}
\def\sigmab{{\bar \sigma}} \def\Sigmab{{\bar \Sigma}}

\def\chib{ {\bar \chi}}

\def\SO{ {\hbox{\rm SO} } }
\def\SOS{ {\hbox{\rm SO}} (7)_{\rm aux} }
\def\OS{ {\hbox{\rm O}} (7)_{\rm aux} }
\def\SU{ {\hbox{\rm SU} } }
\def\U{ {\hbox{\rm U}} }
\def\GT{ {\hbox{\rm G}}_2 }

\def\i{ {\hbox{\bf 1}} }
\def\ii{ {\hbox{\bf 2}} }         \def\iib{ \bar {\hbox{\bf 2}} }
\def\iii{ {\hbox{\bf 3}} }        \def\iiib{ \bar {\hbox{\bf 3}} }
\def\iv{ {\hbox{\bf 4}} }         \def\ivb{ \bar {\hbox{\bf 4}} }
\def\vi{ {\hbox{\bf 6}} }
\def\vii{ {\hbox{\bf 7}} }
\def\viii{ {\hbox{\bf 8}} }
\def\x{ {\hbox{\bf 10}} }
\def\xvi{ {\hbox{\bf 16}} }


\def\npb{Nucl.~Phys.~}
\def\plb{Phys.~Lett.~}

\def\cqg{Class.~Quantum Grav.~}

\def\cmp{Commun.~Math.~Phys.~}
\def\jmp{J.~Math.~Phys.~}


\nopagenumbers{\baselineskip=12pt
\line{\hfill CERN-TH.7231/94}
\line{\hfill hep-th/9404190}
\ifdoublepage \bjump\bjump\bjump\bjump\else\vfill\fi
\centerline{\capsone SUPERSYMMETRY ALGEBRAS}
\sjump
\centerline{\capsone AND LORENTZ INVARIANCE}
\sjump
\centerline{ {\capsone FOR~} {\mtitle d}~{\capsone ~= 10 SUPER YANG-MILLS} }
\bjump\bjump
\centerline{\scaps Jonathan M.~Evans\footnote{$^1$}{
Supported by a fellowship from the EU Human Capital and Mobility programme}}
\sjump
\centerline{\sl Theoretical Physics Division}
\centerline{\sl CERN}
\centerline{\sl CH-1211 Gen\`eve 23}
\centerline{\sl Switzerland}
\sjump
\centerline{ {\sl e-mail:} evansjm@surya11.cern.ch}
\vfill
\ifnum\unoduecol=2 \eject\null\vfill\fi
\centerline{\capsone ABSTRACT}
\sjump
\noindent
We consider ways in which conventional supersymmetry can be embedded
in the set of more general fermionic transformations proposed
recently [\Ref{B}] as a framework in which to study $d=10$ super Yang-Mills.
Solutions are exhibited which
involve closed algebras of various numbers of supersymmetries together with
their invariance groups:
nine supersymmetries with $\GT {\times}\SO (1,1)$ invariance;
eight supersymmetries with $\SO (7){\times}\SO (1,1)$ invariance;
four supersymmetries with $\SO (3,1){\times}\U (3)$ invariance.
We recover in this manner all previously known ways of adding finite numbers
of bosonic auxiliary fields so as to partially close the $d=10$ superalgebra.
A crucial feature of these solutions is that the auxiliary fields
transform non-trivially under the residual Lorentz symmetry,
even though they are originally introduced as Lorentz scalars.
\bjump
\line{CERN-TH.7231/94 \hfill}
\line{April 1994 \hfill}
\sjump
\ifnum\unoduecol=2 \vfill\fi
\eject
\yespagenumbers\pageno=1

\chapter{Introduction and overview}

\noindent
Supersymmetric Yang-Mills theories possess a wealth of fascinating physical
and mathematical properties which have been extensively studied in the last
twenty years.\footnote{$^2$}{General introductions with comprehensive
citations can be found in [\Ref{S},\Ref{GSW}].}
An infamous outstanding problem is the construction of
an off-shell formulation of ten-dimensional super
Yang-Mills.\footnote{$^3$}{The maximally-extended super Yang-Mills theories
were formulated in [\Ref{BSS}]. The auxiliary field problem has been
studied using: component fields [\Ref{SR}]; conventional superspace in
$d{=}4$ [\Ref{D4},\Ref{HST1}], $d{=}6$ [\Ref{HST2}],
and $d{ = }10$ [\Ref{D10}]; light-cone superspace
[\Ref{BLNM},\Ref{BGS},\Ref{J2}]; and harmonic superspace [\Ref{HS}].}
A solution would be of considerable interest
in its own right as well as being likely to offer a new perspective on
other difficult and long-standing questions such as the existence of
covariant quantum actions for superparticles and superstrings in
ten dimensions.\footnote{$^4$}{See [\Ref{GSW}] and, for a more detailed
and up-to-date account, [\Ref{H}].}
Here we investigate a novel setting for this
problem which was suggested recently in [\Ref{B}]. In this
first section we clarify some general aspects
of this approach and formulate our aims.

The Lagrangian for $d = 10$ super Yang-Mills can be written
$$
L = {\rm Tr} \left [ -\frac14 F^{\mu \nu} F_{\mu \nu }
+ \frac{i}2 \psi \G^\mu D_\mu \psi
+ \frac12 G_i G_i
\right ]
\, .
\nfr{lag}
$A_\mu$ is a Yang-Mills gauge field,
$D_\mu = \del_\mu + A_\mu$ is the associated covariant derivative,
$F_{\mu \nu}= \del_\mu A_\nu - \del_\nu A_\mu + [A_\mu , A_\nu]$
is the field strength and $\psi$ is a sixteen-component Majorana-Weyl spinor.
The fields $A_\mu$ and $\psi$ describe
equal numbers of propagating {\it on\/}-shell modes and the
auxiliary scalar fields $G_i$ with $i = 1 , \ldots , 7$
are included to balance the bosonic and fermionic
{\it off\/}-shell degrees of freedom.
All fields takes values in the Lie algebra of some
gauge group, `Tr' denotes an invariant inner-product on this algebra
and $L$ is, of course, gauge-invariant.
The classical gauge coupling constant has been scaled out of the
Lagrangian.

The Lagrangian $L$ is invariant under the transformations
$$\eqalign{
\d_{(\e , v_i)} A_\mu & = i \e \G_\mu \psi \cr
\d_{(\e , v_i)} \psi & = \frac12 F^{\mu \nu} \G_{\mu \nu} \e + G_i v_i \cr
\d_{(\e , v_i)} G_i & = - i v_i \G^\mu D_\mu \psi \cr
}\nfr{gsusy}
which depend not only on the usual spinor parameter $\e$ but
also on an additional seven spinor parameters $v_i$ satisfying the conditions
$$
v_i \G_\mu \e = 0
\, , \qquad
v_i \G_\mu v_j = \delta_{ij} \e \G_\mu \e
\, .
\nfr{veqns}
The parameters $\e$ and $v_i$ are taken to be
{\it commuting\/} spinors, which means that the variation $\d_{(\e , v_i)}$
is fermionic in character, but spinor fields such as $\psi$ must be treated
as {\it anticommuting\/}.
We shall refer to a transformation \gsusy\ with the additional
restriction \veqns\ as a {\it generalized supersymmetry\/}.
It can be shown by direct computation that any
two generalized supersymmetries
obey, up to field equations,
the standard supersymmetry algebra
$$
\{ \d_{(\e , v_i)} , \d_{(\eh , \vh_i)}  \} =
- 2 i \e \Gamma^\mu \eh \, D_\mu
\nfr{calg}
when acting on $D_\mu$, $\psi$ or $G_i$ (considering the action on $D_\mu$
rather than $A_\mu$ just gives a neat way of including the relevant gauge
transformations in the algebra) and that
this algebra holds independently of field equations when
$(\e , v_i) = (\eh , \vh_i)$.
One can use this fact to find
larger sets of generalized supersymmetries obeying a closed algebra
off shell and it was argued in [\Ref{B}] that a
maximum of nine generalized supersymmetries can be found with this property.

The attractive feature of this framework is that the
Lagrangian \lag , the transformations \gsusy\
and the additional constraints \veqns\ are manifestly
invariant under the Lorentz group SO(9,1).
They are also clearly invariant with respect to an internal symmetry group
$\OS$ under which the auxiliary fields $G_i$ and the spinor parameters $v_i$
transform as seven-dimensional vectors. The suffix
`aux' indicates that these latter transformations do not alter
physical states of the theory; nevertheless they have an important role to
play. For one thing the equations \veqns\ are sufficient to determine
the spinors $v_i$ from a given $\e$ precisely up to a transformation
of type $\OS$.
We shall be concerned with the connected invariance group
$\SO (9,1) \times \SOS$ of the complete set of generalized
supersymmetries, and with the fact that this cannot, unfortunately,
be preserved in choosing subsets of supersymmetries
which obey the algebra (1.4) off shell.

Our aim is to study ways in
which conventional supersymmetry transformations can be
embedded within the set of generalized supersymmetries.
Conventional supersymmetries
depend linearly on the single spinor parameter $\e$
and so our task is to find solutions of \veqns\ of the form
$$
v_i = M_i \e
\nfr{redsol}
for some suitable matrices $M_i$.
Any such choice must break Lorentz invariance because in $d=10$ there are
no non-trivial Lorentz-invariant tensors of the required type.
In order to construct solutions
we shall find it necessary to restrict the spinor $\e$ to some subspace.
The choice of this subspace together with the choice of matrices $M_i$
will determine a particular subgroup of $\SO (9,1) \times \SOS$
which survives as the invariance group of the solution.
In sections 2, 3 and 4 below we present solutions involving
various numbers of supersymmetries together with detailed discussions
of their residual invariance groups and associated representations.

One subtle aspect of the residual symmetry is worth pointing out in advance.
In each of the cases below we shall find the pattern of symmetry
breaking\footnote{$^5$}{
Arrows will always indicate reductions in symmetry groups and decompositions
of representations, {\it never\/} homomorphisms between groups.}
$\SO (9,1) \times \SOS \to \SO (n,1) \times H$, for some $n$ and some group
$H$. But here $H$ is embedded diagonally in the factors on the
left-hand side: the Lorentz factor is broken
$\SO (9,1) \to \SO (n,1) \times H$
and the internal auxiliary symmetry is broken
$\SOS \to H$ in such a way that these copies of the group $H$ get identified.
Since $\SOS$ has no direct
physical significance we are entitled to interpret the surviving
symmetries belonging to $H$ as Lorentz transformations. But
the auxiliary fields transform non-trivially under $\SOS$ and hence under $H$.
The auxiliary fields in our solutions
will therefore behave non-trivially under residual
Lorentz transformations even though they are originally
introduced as $\SO (9,1)$ scalars.

The most important property of our solutions is that the algebra
\calg\ will hold off-shell.
We mentioned above that \calg\ holds
when $( \e , v_i ) = ( \eh , \vh_i )$
and that in [\Ref{B}] a method was given to generate
other generalized supersymmetries which all obey a closed algebra.
This method is more general than we need here, however, because we are
considering solutions of \veqns\ which are all related by
\redsol\ with the spinor parameter $\e$ restricted to some particular
subspace. In these circumstances the
corresponding generalized supersymmetries depend linearly on
$\e$ and, denoting them simply by
$\d_{\e} = \d_{(\e , M_i \e)} $, it follows that any two
such transformations obey
$\{ \d_{\e} , \d_{\eh} \} = \frac12 ( \, \{ \d_{\e + \eh} , \d_{\e + \eh} \}
- \{\d_{\e} , \d_{\e} \} - \{ \d_{\eh} , \d_{\eh} \} \, ) $.
We know that \calg\ holds for each term on the right-hand side and
it follows that it must hold for the expression on the
left-hand side too.
Hence, any generalized supersymmetries corresponding to a particular space of
solutions of \veqns\ and \redsol\ will automatically obey a closed algebra.

It is natural to ask how our results compare with previous work
in which auxiliary fields have been found for subsets of $d=10$ supersymmetry
transformations.
One possibility is to select just those
supersymmetries which preserve a light-cone gauge condition
[\Ref{BLNM},\Ref{BGS},\Ref{J2}].
In [\Ref{J2}]
a systematic search was made for bosonic auxiliary fields which would close
such a light-cone supersymmetry algebra;
it was shown that there is no solution which is covariant with respect to
the residual SO(8) part of the light-cone symmetry group,
but that a solution exists which is covariant with respect to
an SO(7) subgroup. The solution of section 2 generalizes this result.
Another option, distinct from the light-cone approach,
is to consider the trivial dimensional reduction of super Yang-Mills in $d=10$
to theories with $N=2$ supersymmetry in $d=6$ or $N=4$ in $d=4$.
It is easy to find an $N=1$ superspace formulation of the theory in $d=4$
by coupling three chiral matter multiplets to $N=1$, $d=4$ Yang-Mills
[\Ref{S}]. This case arises as a corollary of the solution given in section 4.
It is also possible, after overcoming some highly
non-trivial obstacles, to find superspace formulations
which make manifest $N=2$ supersymmetry in $d=4$ [\Ref{HST1}] or $N=1$
supersymmetry in $d=6$ [\Ref{HST2}]. These last possibilities involve
fermionic auxiliary fields, however, and so one should not expect to find them
appearing in the framework considered here.
Finally, in [\Ref{B}] a particular solution to \veqns\ was presented using
octonionic notation. We recover this in section 3 in conventional notation
by modifying the solution of section 2.

The last chore before giving the solutions is to fix some details of
notation. Spacetime indices in $d = 10$ are written
$\mu , \nu = 0 , \ldots , 9$ and they label the vector representation
$\x$ of SO(9,1). The Minkowski metric is defined by
$-\eta_{00} = \eta_{11} = \ldots = \eta_{99} = 1$.
Spinor indices will often be suppressed but when they are needed
upper and lower indices
$\A, \B = 1 , \ldots , 16$ will denote the inequivalent Majorana-Weyl
representations $\xvi_{\pm}$. By definition $\psi$, $\e$ and $v_i$ all belong
to the $\xvi_+$ representation.
Upper and lower spinor indices cannot be lowered or raised but they can be
contracted invariantly.
The corresponding gamma matrices $(\Gamma_\mu)^{\A \B}$ and
$(\Gamma_\mu)_{\A \B} $ are symmetric and obey
$
(\G_\mu)^{\A \C} (\G_\nu)_{\C \B} +
(\G_\nu)^{\A \C} (\G_\mu)_{\C \B} =
2 \eta_{\mu \nu} \d^\A{}_\B$.
Antisymmetrized products are denoted in the
usual ways, {\it eg.\/}~$
(\G_{\mu \nu})^\A{}_\B = \frac12 [ \,
(\G_{\mu})^{\A \C} (\G_{\nu})_{\C \B}^{\phantom{A}} -
(\G_{\nu})^{\A \C} (\G_{\mu})_{\C \B}^{\phantom{A}} \,
]$
is the generator of Lorentz transformations in the $\xvi_+$
representation.
Because the spinor representations are Majorana, it is always possible
to choose a basis in which all components of spinors and gamma matrices are
real. Because the spinor indices label Weyl representations, we can also take
the gamma matrices to obey $(\G_{0\ldots9})^A{}_B = \d^A{}_B$.
\vfill \eject

\chapter{Eight supersymmetries with SO(7)$\times$SO(1,1) invariance}
\noindent
We start by considering a decomposition of the Lorentz group to a
light-cone subgroup
$
\SO (9,1) \rightarrow \SO (8) \times \SO (1,1)
$
in which the first factor acts on the subspace of Minkowski space
with coordinates labeled by $\mu = 1, \ldots , 8$
and the second factor acts on the subspace with coordinates
labeled by $\mu = 0, 9$.
With respect to this subgroup the vector
representation decomposes
$\x \to \viii^0 \oplus \i^{2} \oplus \i^{-2}$
and the spinor representations decompose according to
$\xvi_+ \to \viii_+^{1} \oplus \viii_-^{-1}$ and
$\xvi_- \to \viii_-^{1} \oplus \viii_+^{-1}$.
Here $\viii$ and $\viii_\pm$ denote the vector and
spinor representations of SO(8) respectively
and superscripts specify the eigenvalue of
a suitably normalized generator of the SO(1,1) factor.
We will be concerned mostly with the decomposition of the $\xvi_+$
representation for which this generator is $\G_{09} = \G_{1 \ldots 8}$.

We will look for a solution of \veqns\ in which the supersymmetry
parameter satisfies the restriction
$$
\G_{09} \e = \e
\nfr{conda}
giving eight linearly-independent supersymmetries.
To obtain such a solution it appears to be necessary to
break the symmetry still further by
selecting a particular transverse direction, which may
as well be the one labeled by $\mu = 8$.
We then define
$$
v_i = \G_{i8} \e \, , \qquad i = 1, \ldots , 7
\nfr{sola}
and claim that \conda\ and \sola\ provide a solution of \veqns .
This can be checked
using standard gamma matrix manipulations; of particular use is the fact that
$\G_{\mu \nu \rho}$ is antisymmetric in its spinor indices so that
$\e \G_{\mu \nu \rho} \e = 0$.

In choosing a particular transverse direction we have clearly compounded
the reduction in symmetry by breaking
$\SO (8) \to \SO (7)$.
Under this subgroup the vector representation of course
decomposes $\viii \to \vii \oplus \i$ while the spinor representations remain
irreducible but become isomorphic:
$\viii_\pm \to \viii$ (it should be obvious from the context
whether $\viii$ denotes the vector representation of SO(8) or the
spinor representation of SO(7)).
In line with the general remarks made in the introduction, we see that
\sola\ identifies this $\SO (7)$ subgroup
with the group $\SOS$, because
the index $i$ on the left-hand side of \sola\ was originally an internal
label for the $\vii$ of $\SOS$ whereas on the right-hand side it labels a
direction in spacetime.
The residual symmetry group is
just the diagonal subgroup of these two SO(7) factors under which
the auxiliary fields $G_i$ transform as a vector.

To summarize, we have found a closed algebra of eight supersymmetries
with the residual invariance group $\SO (7) \times \SO (1,1)$
and representations
$$\eqalign{
A_\mu & : \, \vii^0 \oplus \i^0 \oplus \i^{2} \oplus \i^{-2} \cr
\psi \, & : \, \viii^{1} \oplus \viii^{-1} \cr
G_i & : \, \vii^0 \cr
\e \, & : \,  \viii^{1} \cr
}
$$

It is instructive to write this solution
explicitly in terms of irreducible representations of the invariance group.
To do so we consider the decomposition of a spinor
$\chi^\A \to (\chi^+_\a , \chi^-_\a )$, where the indices
$\a , \b = 1 , \ldots , 8$
label the spinor representation of SO(7) and the superscripts indicate the
eigenvalue of $\G_{09}$. A suitable
corresponding block form for the $d =10$ gamma matrices is given by
$$\eqalign{
-&(\G_0)_{\A \B} = (\G_0){}^{\A \B} =
\pmatrix{
1 & 0 \cr
0 & 1 \cr
}
\qquad
(\G_9)_{\A \B} = (\G_9){}^{\A \B} =
\pmatrix{
1 & 0 \cr
0 &-1 \cr
}
\cr
&(\G_8)_{\A \B} = (\G_{8})^{\A \B} =
\pmatrix{
0 & 1 \cr
1 & 0 \cr
}
\qquad
(\G_i)_{\A \B} = (\G_i)^{\A \B} =
\pmatrix{
0 & \lambda_i \cr
-\lambda_i & 0 \cr
}
\cr
}
\nfr{glcone}
where $i = 1, \ldots , 7$.
The matrices $(\lambda_i)_{\a \b}$ are real and antisymmetric and
they obey
$$
\lambda_i \lambda_j + \lambda_j \lambda_i = - 2 \delta_{ij}
\nfr{scliff}
(an explicit construction of these matrices can be found in the appendix).
Denoting antisymmetrized products in the usual way, the matrices
$(\lambda_{ij})_{\a \b}$ are antisymmetric and generate the
spinor representation of SO(7).
The solution \conda , \sola\ can now be written simply as
$$
v^+_i = \lambda_i \e^+ \, , \qquad v^-_i = \e^- = 0 \, .
\nfr{solb}
If we choose
the light-cone gauge $A_0 {+} A_9 = 0$ we recover
exactly the construction introduced previously in [\Ref{J2}].
In addition to understanding how this construction fits into the
framework of generalized supersymmetry, we
have now learned that such a light-cone gauge choice
is not necessary in order to close the light-cone supersymmetry algebra.

As a last point of interest, we note that one can use triality,
{\it ie.\/}~the $S_3$ group of outer automorphisms of SO(8), to construct
similar solutions invariant under inequivalent SO(7) subgroups.
There are essentially three inequivalent SO(7) subgroups of SO(8) whose vector
representations sit in the `obvious' ways inside the vector or spinor
representations of SO(8) -- see {\it eg.\/}~[\Ref{A},\Ref{P}].
The construction of solutions
invariant under these alternative SO(7) subgroups is essentially similar to
the solution presented above and we omit the details.

\chapter{Nine supersymmetries with G${}_2{\times}$SO(1,1) invariance}
\noindent
We now show how a similar solution to that of the
last section can be `enlarged' by
one more supersymmetry, but only at the expense of
reducing the invariance group.
In the notation introduced in the last section,
we seek a solution in which $\e^-$ is non-zero.
We pick out a particular direction in spinor space,
which may as well be the one labeled by $\a = 8$, and
define a diagonal matrix $n_{\a \b}$ by
$-n_{11} = \ldots = -n_{77} = n_{88} = 1$.
We claim that \veqns\ is satisfied if
$$
v_i^+ = n \lambda_i n \e^+ \, ,
\qquad
v_i^- = - \lambda_i \e^-  \, ,
\qquad
n \e^- = \e^- \, .
\nfr{solc}
To check this it is convenient to introduce the
matrix $p = \frac12 (1 + n)$ which projects onto the one-dimensional
positive eigenspace of $n$.
The solution can then be verified by
direct substitution using the expressions in
\glcone\ and using also the facts
$p \lambda_i p =0$, $p \lambda_i \lambda_j p = - \delta_{ij} p$
and $p \lambda_i \lambda_j \lambda_k p = p \lambda_{ijk} p$
which are all simple consequences of \scliff .

Since $\e^+$ is arbitrary but $n\e^- = \e^-$ we have found
a solution with nine independent supersymmetries.
It can be shown that this coincides with
the solution presented in [\Ref{B}] using octonionic notation but we shall
not give the details here; it can be demonstrated in
a pedestrian way by writing out in conventional notation the results of all
octonionic multiplications in the expressions in [\Ref{B}].
Now we consider the invariance properties of this solution, which
were not dealt with in [\Ref{B}]. The residual invariance clearly
consists of an SO(1,1) factor together with the subgroup of SO(7) which
fixes the particular direction $\a = 8$ in the spinor representation,
which is equivalent to fixing the matrix $n_{\a \b}$.
It is well-known [\Ref{G},\Ref{P}] that this subgroup is $\GT$ and that one
can write down explicit expressions for its generators.
The combination of SO(7) generators $a_{ij} \lambda_{ij}$
leaves the matrix $n$ inert precisely when $c_{ijk} a_{jk} = 0$, where
$c_{ijk} = (\lambda_i)_{jk}$ is completely antisymmetric.
These seven conditions on the twenty-one
generators of SO(7) leave fourteen independent combinations,
as required for $\GT$.
The vector and spinor representations decompose under
$ \SO (7) \rightarrow \GT $ according to
$\vii \to \vii$ and
$ \viii \to \vii \oplus \i$ respectively.

By combining these results with those of the last section
we can read off the final transformation properties of all the fields.
We have found a solution giving a closed algebra of
nine supersymmetries with invariance group $\GT \times \SO (1,1)$
and representations
$$\eqalign{
A_\mu &: \, \vii^0 \oplus \i^0 \oplus \i^{2} \oplus \i^{-2} \cr
\psi \, &: \, \vii^1 \oplus \i^1 \oplus \vii^{-1} \oplus \i^{-1} \cr
G_i &: \, \, \vii^0 \cr
\e \, &: \, \vii^{1} \oplus \i^{1} \oplus \i^{-1} \cr
}
$$

\chapter{Four supersymmetries with SO(3,1)$\times$U(3) invariance}
\noindent
We start from the decomposition of the Lorentz group
$\SO (9,1) \to \SO (3,1) \times \SU (4)$
where the first factor acts on the subspace with coordinates labeled
by $\mu = 0, 1, 2, 3$ and the second factor acts on the subspace
labeled by $\mu = 4, \ldots , 9$.
The vector representation decomposes
$\x \to (\iv , \i ) \oplus (\i , \vi )$ and the spinors
decompose according to
$\xvi_+ \to (\ii , \iv) \oplus (\iib , \ivb )$ and
$\xvi_- \to (\ii , \ivb ) \oplus (\iib , \iv )$.
We will be concerned mostly with the representation $\xvi_+$
for which the irreducible subspaces are
just the eigenspaces
of the matrix $ - \G_{0123} = \G_{456789}$
with eigenvalues $\pm i$.

As in section 2, we will present the solution first using $d=10$ notation.
We demand that the supersymmetry parameter
satisfies
$$
\G_{45} \e = \G_{67} \e = \G_{89} \e \, .
\nfr{condd}
This amounts to two linearly independent conditions, each of which
halves the dimension of the spinor, leaving us with four supersymmetries.
We then define
$$\eqalign{
v_1 & = \G_{68} \e = - \G_{79} \e \, , \qquad
v_2 
= - \G_{69} \e = - \G_{78} \e \, , \cr
v_3 & = \G_{84} \e = - \G_{95} \e \, , \qquad
v_4 
= - \G_{85} \e = - \G_{94} \e \, , \cr
v_5 & = \G_{46} \e = - \G_{57} \e \, , \qquad
v_6 
= - \G_{47} \e = - \G_{56} \e \, , \cr
& \qquad v_7
= - \G_{45} \e = - \G_{67} \e = - \G_{89} \e \, , \cr
}
\nfr{sold}
and claim that this is a solution of \veqns .
These alternative expressions, which all follow
from the condition \condd\ on $\e$, enable one to check this claim quite
quickly.
The idea is that by selecting particular expressions for each of the
$v_i$ from the possibilities given above
one can easily see that all undesired terms
in \veqns\ take the form $\e \G_{\mu \nu \rho} \e = 0$. (We noted earlier
that this combination of gamma matrices is always antisymmetric in its spinor
indices.)

The solution \sold\ is clearly SO(3,1) covariant but it is less obvious
that the surviving subgroup of the SU(4) factor is U(3).
It is possible, with some effort, to demonstrate this using $d =10$ notation
but we choose instead to expose this symmetry by
passing to a complex basis in which spinors take the form
$\chi^\A \to ( \chi^{\a a} , \chib_{\ad a})$ and $\chi_\A \to
( \chi_{\a a} , \chib^{\ad a} )$.
Here $\a , \ad = 1, 2$ label the $\ii$ and $\iib$ of SO(3,1) and these
indices can be raised and lowered according to
$\chi^\a = \ve^{\a \b} \chi_\b$, $\chi_\b = \chi^\a \ve_{\a \b}$,
$\chib^\ad = \ve^{\ad \bd} \chib_\bd$, $\chib_{\bd} = \chib^{\ad} \ve_{\ad \bd}
$ where the antisymmetric symbols are defined by $\ve^{12} = \ve_{12} = 1$.
The upper and lower indices $a = 1, 2, 3, 4$ label the $\iv$ and $\ivb$
of SU(4) and cannot be lowered or raised.
In such a basis the gamma matrices have the block structure
$$\eqalign{
&(\G_{\mu})^{\A \B}
= \pmatrix{
0 &
(\sigma_\mu)^\a{}_\bd \delta^a{}_b \cr
(\sigmab_\mu)_\ad{}^\b \delta_a{}^b
& 0 \cr
}
\qquad
(\G_{m+3})^{\A \B}
=\pmatrix{
\ve^{\a \b} (\Sigma_m)^{ab} & 0 \cr
0 & \ve_{\ad \bd} (\Sigmab_m)_{ab} \cr
} \cr
&(\G_{\mu})_{\A \B}
= \pmatrix{
0 &
(\sigma_\mu)_\a{}^\bd \delta_a{}^b \cr
(\sigmab_\mu)^\ad{}_\b \delta^a{}_b
& 0\cr
}
\qquad
(\G_{m+3})_{\A \B}
=\pmatrix{
\ve_{\a \b} (\Sigmab_m)_{ab} & 0 \cr
0 & \ve^{\ad \bd} (\Sigma_m)^{ab} \cr
} \cr
}
\nfr{gufic}
where $\mu = 0, 1, 2, 3$ and $m = 1, \ldots , 6$.
The matrices $(\sigma_\mu)^{\a \bd}$ and $(\sigmab_\mu)_{\ad \b}$
obey $\sigma_\mu \sigmab_\nu + \sigma_\nu \sigmab_\mu = - 2 \eta_{\mu \nu}$
and $(\sigma_\mu)^\a{}_\bd = (\sigmab_\mu)_\bd{}^\a$.
The matrices $(\Sigma_m)^{ab}$ and $(\Sigmab_m)_{ab}$ are antisymmetric,
they obey $\Sigma_m \Sigmab_n + \Sigma_n \Sigmab_m = - 2 \delta_{mn}$,
and they can be taken to be complex conjugates of one another.
Explicit constructions of both sets of matrices are given in the appendix.

Armed with this notation the condition \condd\ can be replaced by
$$
\e^{\a a} = \eb_{\ad a} = 0 \, , \qquad a = 1, 2, 3
\, .
\nfr{conde}
It is convenient to drop the label $a = 4$ on the only non-zero components of
$\e$, writing these simply as $\e^\a$ and $\eb_\ad$.
Now \sold\ can be replaced by the expressions
$$\eqalign{
(v_m)^{\a a} = (\Sigma_m)^{a4} \e^\a \, &, \qquad
(\vb_m)_{\ad a} = (\Sigmab_m)_{a4} \eb_{\ad} \, , \cr
(v_7)^{\a 4} = i \e^\a \, &, \qquad (\vb_7)_{\ad 4} = - i \eb_{\ad} \, , \cr
}
\nfr{sole}
with $m =1 , \ldots , 6$ and $a = 1, 2, 3$ and all other components zero.
The alternative forms of the solution \condd , \sold\ and \conde , \sole\
can be shown to coincide using \gufic\ and the explicit expressions for
$\Sigma_m$ and $\Sigmab_m$ given in the appendix.
It can also be checked directly that \conde\ and \sole\ provide
a solution of \veqns .

The condition \conde\ breaks the Lorentz factor
$\SU (4) \to \U (3) = \SU (3) {\times} \U (1)$ which is then
preserved by the full solution \sole .
There is a similar pattern of
breaking for the auxiliary internal symmetry
$\SOS \to \SU (4) \to \SU (3) \times \U(1)$ (where the first reduction
occurs as a result of picking out the direction $i =7$). Once again
the surviving subgroups of the Lorentz and auxiliary symmetries are
identified by the solution. For the reduction
$\SU (4) \to \SU (3)$ the quantities
$(\Sigma_m)^{a4}$ and $(\Sigmab_m)_{a4}$ are exactly the invariant tensors
which describe the decomposition $\vi \to \iii \oplus \iiib$,
and so the $\SU (3)$ factor of the final invariance is manifest in \sole .
The $\U (1)$ factor is more subtle because it involves a non-trivial
relative normalization between the two groups; we simply state below the final
result for the various U(1) weights.

We have found a closed algebra of four supersymmetries
with invariance group $\SO (3,1) \times \SU (3) \times \U (1)$ and
representations
$$\eqalign{
A_\mu &: \, (\iv , \i)^0 \oplus (\i , \iii )^{-2} \oplus (\i , \iiib)^{2}\cr
\psi \, &: \, (\ii , \i )^{-3} \oplus (\iib , \i )^{3} \oplus
(\ii , \iii )^{1} \oplus (\iib , \iiib )^{-1} \cr
G_i &: \, (\i , \i )^0 \oplus (\i , \iii )^{4} \oplus (\i , \iiib )^{-4} \cr
\e \, &: \, (\ii , \i )^{-3} \oplus (\iib , \i)^{3} \cr
}
$$

Given the residual symmetry group of this solution, it is natural to ask
what happens if we perform a trivial dimensional reduction from $d=10$ to
$d=4$. The answer is that we obtain
off-shell $N=1$ Yang-Mills in the Wess-Zumino gauge
coupled to three chiral $N=1$ matter multiplets [\Ref{S}].
The components of the former can be taken to be
$A_\mu$ $(\mu = 0, 1, 2, 3)$, $\psi^{\a 4}$, $G_7$,
while the components of the latter can be taken to be
$\phi^a = (\Sigma_m)^{a4} A_{m + 3}$, $\psi^{\a a}$,
$K^a = (\Sigma_m)^{a4} G_m$ $(a= 1, 2, 3)$.
The fields $K^a$ are not quite the usual matter auxiliary fields
(because their equations of motion are $K^a =0$) but they are
related to them in a simple way.
In $d=4$ the residual $\SU (3) \times \U (1)$ invariance becomes an
internal symmetry with a rather nice interpretation.
The SU(3) factor acts on the three chiral multiplets
in an obvious way; the U(1) factor is an example of an $R$-symmetry
and it can be checked that the weights given above can be obtained
by applying the general prescription of [\Ref{GGRS}] to this model.

\chapter{Concluding remarks}
\noindent
In this paper we have clarified how off-shell algebras of
conventional supersymmetries can exist within the framework of generalized
supersymmetry. In doing so we have recovered
all previously known ways of adding finite numbers
of bosonic auxiliary fields so as to partially close the $d=10$
superalgebra.
The auxiliary fields in these solutions must eventually transform
non-trivially under remnants of the $d=10$ Lorentz group, even though
they are introduced as SO(9,1) scalars.
We have seen in each case how this is made possible by
the existence of the internal auxiliary symmetry $\SOS$.

Our results lend further weight to the
idea of generalized supersymmetry introduced in [\Ref{B}] and
provide strong motivation for its future study.
It would be interesting to try and include non-propagating fermionic
degrees of freedom so as to
reproduce the more complicated sets of auxiliary fields
given in [\Ref{HST1},\Ref{HST2},\Ref{J2}].
It would also be very interesting to find some superspace description
of generalized supersymmetry transformations with the exciting possibility
that this might lead to new covariant actions for superparticles and
superstrings.
\bjump

\centerline{\capsone ACKNOWLEDGMENTS}
\sjump
\noindent
I am grateful to Nathan Berkovits for conversations and correspondence
concerning his work and to Hugh Osborn for first bringing my attention
to [\Ref{B}].

\vfill \eject

\centerline{\capsone APPENDIX}
\sjump
\noindent
Let $\t_1$, $\t_2$, $\t_3$ be the usual Pauli matrices.
For the matrices $\lambda_i$ of section 2 we can take
$$\eqalign{
&
\lambda_1 = i\t_2 \otimes i\t_2 \otimes i\t_2 \, , \quad
\lambda_2 = \t_1 \otimes i\t_2 \otimes 1 \, , \quad
\lambda_3 = i\t_2 \otimes 1 \otimes \t_1 \, , \quad
\lambda_4 = -i \t_2 \otimes 1 \otimes \t_3 \, ,
\cr
&
\qquad \quad \lambda_5 = 1 \otimes \t_1 \otimes i \t_2 \, , \qquad
\lambda_6 = - \t_3 \otimes i \t_2 \otimes 1 \, , \qquad
\lambda_7 = - 1 \otimes \t_3 \otimes i \t_2 \, .
\cr
}$$
For the matrices $(\sigma_\mu)^{\a \ad}$ and $(\sigmab_\mu)_{\ad \a}$
of section 4 we can take
$$
\sigma^0 = \sigmab^0 = 1 \, , \qquad
\sigma_a = - \sigmab_a = \t_a \, , \quad a= 1,2,3 \, .
$$
For the matrices $(\Sigma_m)^{\A \B}$ and $(\Sigmab_m)_{\A \B}$ of section 4
we can take
$$\eqalign{
&
\Sigma_1 = i \t_2 \otimes \t_1 \, , \qquad
\Sigma_2 = - \t_1 \otimes \t_2 \, , \qquad
\Sigma_3= - i \t_2 \otimes \t_3 \, ,
\cr
&
\Sigma_4 = - \t_2 \otimes 1 \, , \qquad
\Sigma_5 = i 1 \otimes \t_2 \, , \qquad \quad
\Sigma_6 = \t_3 \otimes \t_2 \, ,
\cr
}$$
with $\Sigmab_m$ the complex conjugate of $\Sigma_m$.
\bjump

\centerline{\capsone REFERENCES}
\sjump
\beginref
\Rref{B}{N.~Berkovits, \plb {\bf B318} (1993) 104}
\Rref{J2}{J.M.~Evans, \npb {\bf B310} (1988) 108}
\Rref{BSS}{L.~Brink, J.~Scherk and J.H.~Schwarz, \npb {\bf B121} (1977) 77;
\hfil \break
F.~Gliozzi, J.~Scherk and D.~Olive, \npb {\bf B122} (1977) 253}
\Rref{BGS}{L.~Brink, M.B.~Green and J.H.~Schwarz, \npb {\bf B223} (1983) 125}
\Rref{BLNM}{L.~Brink, O.~Lindgren and B.E.W.~Nilsson, \npb {\bf B212} (1983)
401; \hfil \break
S.~Mandelstam, \npb {\bf B213} (1983) 149}
\Rref{GSW}{M.B.~Green, J.H.~Schwarz and E.~Witten,
{\it Superstring theory volume 1: introduction\/} (CUP, 1987)}
\Rref{S}{M.~Sohnius, Phys.~Rep.~{\bf 128} (1985) 39}
\Rref{HST1}{P.~Howe, K.S.~Stelle and P.K.~Townsend, \npb {\bf B214} (1983) 519}
\Rref{HST2}{P.~Howe, G.~Sierra and P.K.~Townsend, \npb {\bf B221} (1983) 331}
\Rref{SR}{W.~Siegel and M.~R\v ocek, \plb {\bf B105} (1981) 275}
\Rref{D4}{M.F.~Sohnius, \npb {\bf B136} (1978) 461;
E.~Witten, \plb {\bf B77} (1978) 394;
W.~Siegel, \plb {\bf B80} (1979) 220;
J.~Harnard, J.~Hurtubise, M.~L\'egar\'e and S.~Shnider, \npb {\bf B256}
(1986) 183
}
\Rref{D10}{W.~Siegel, \plb {\bf B80} (1979) 220;
B.E.W.~Nilsson, G\"oteborg preprint 81-6 (1981) unpublished;
E.~Witten, \npb {\bf B266} (1986) 245;
J.~Harnard and S.~Shnider, \cmp {\bf 106} (1986) 183;
E.~Abdalla, M.~Forger and M.~Jacques, \npb {\bf B307} (1988) 198
}
\Rref{HS}{A.~Galperin, E.~Ivanov, S.~Kalitzin, V.~Ogievetski and E.~Sokatchev,
\cqg {\bf 2} (1985) 155;
E.~Sokatchev, \plb {\bf B169} (1986) 209;
E.~Nissimov, S.~Pacheva and S.~Solomon, \npb {\bf B299} (1988) 183;
E.~Nissimov, S.~Pacheva and S.~Solomon, \npb {\bf B317} (1989) 344
}
\Rref{H}{C.M.~Hull and J-L.~V\'azquez-Bello, Queen Mary and Westfield College
preprint QMW-93-07, hep-th/9308022}
\Rref{G}{M.~G\"unaydin and F.~G\"ursey, \jmp {\bf 14} (1973) 1651;
R.~D\"undarer, F.~G\"ursey and C-H.~Tze, \jmp {\bf 25} (1984) 1496;
E.~Corrigan, C.~Devchand, D.B.~Fairlie and J.~Nuyts, \npb {\bf B214} (1983)
452}
\Rref{GGRS}{S.J.~Gates, M.T.~Grisaru, M.~R\v ocek and W.~Siegel,
{\it Superspace\/} (Benjamin Cummings, 1983) p152}
\Rref{A}{J.F.~Adams, {\it Spin(8), triality}, $F_4$ {\it and all that\/},
in {\it Superspace and Supergravity\/}, edited by S.W.~Hawking and M.~R\v ocek
(CUP, 1981) 435}
\Rref{P}{I.~Porteus, {\it Topological geometry (2nd edition)\/} (CUP, 1982)}
\endref

\ciao